\documentclass[prb,twocolumn,showpacs,preprintnumbers,amsmath,amssymb]{revtex4}
\usepackage{graphicx}
\begin{document}

\preprint{APS/123-QED}
\title {Inverse freezing in the Hopfield Fermionic Ising Spin Glass}
\author{S. G. Magalhaes}%
\email{ggarcia@ccne.ufsm.br}
\affiliation{Departamento de Fisica, Universidade Federal de Santa Maria,
97105-900 Santa Maria, RS, Brazil}%
\author{F. M. Zimmer}
\affiliation{Departamento de F\'\i sica, Universidade do Estado de Santa Catarina, 89223-100, Joinville, SC, Brazil
}
\author{C. V. Morais}
\affiliation{Departamento de Fisica, Universidade Federal de Santa Maria,
97105-900 Santa Maria, RS, Brazil}
\date{\today}

\begin{abstract}

In this work it is studied
the Hopfield fermionic spin glass model
which allows 
  interpolating 
from  trivial randomness to a highly frustrated regime. 
Therefore, it is possible to 
investigate  whether or not   
frustration
is 
 an 
essential
ingredient 
which 
would allow this
magnetic  disordered model 
to present 
naturally inverse freezing by comparing the two limits, 
trivial randomness and highly frustrated regime
and 
how different levels of frustration could affect such unconventional phase transition.
The problem is expressed in the path integral formalism where the spin operators are represented by bilinear combinations of Grassmann variables. The Grand Canonical Potential is obtained within the static approximation and 
one-step replica symmetry breaking scheme.
As a result, phase diagrams temperature  {\it versus} the chemical potential 
are obtained for several levels of frustration.  
Particularly, 
when the level of
frustration is diminished,
the reentrance related  to the inverse freezing is gradually suppressed.

\end{abstract}


\maketitle
\section{Introduction}

There is 
current 
interest  (see, for instance, Refs.  \onlinecite{Schupper1,Crisanti,Sellito,Prestipino,Leuzzi,Angelini,Magal1}) 
in
studying 
inverse 
transitions, melting or freezing,
motivated by 
the 
quite 
unconventional 
situation present in such transitions
in which 
the ordered phase is more entropic than the disordered one \cite{Greer}.
This apparent 
counter-intuitive 
transition
has become even more interesting since there are now various 
physical systems displaying 
this kind of transition, 
as for example, magnetic films \cite{Thin} and, particularly, high-$T_{c}$ superconductors \cite{Super}.
Therefore, the knowledge 
about which 
conditions 
are necessary for the
existence of 
inverse transitions 
is
a challenging issue \cite{Debenedetti}.

Among the proposed classical  magnetic models which can show inverse  transitions,
the Blume Capel (BC) \cite{Schupper1} 
and the Gathak-Sherrington (GS) \cite{Schupper1,Crisanti,Leuzzi} models 
can be 
useful 
to clarify what would be  
such  conditions. 
For instance, the 
phase diagram of the BC model \cite{BC}
displays
inverse melting with
a phase transition 
from the 
ferromagnetic ($FM$) phase to the paramagnetic ($PM$) one when the temperature is decreased.
However,
in order to obtain inverse melting  in the BC model,
it is strictly necessary to impose in the problem
the entropic advantage of the interacting state  
through
parameter $r=l/k \geq 1$,
where $l$
and $k$ are the degeneracy of $\pm 1$ 
and $0$ spins states, respectively
\cite{Schupper1}.
In contrast 
with 
the BC model, 
the 
GS model \cite{GS} - which 
is 
similar to the
the BC model
except that it
has 
the spins couplings 
given
as random Gaussian variables -  
shows  naturally inverse freezing.
In other words, there is a reentrant first order boundary phase separating the spin glass ($SG$) and $PM$ phases. 
However, there is 
no need of any 
entropic 
advantage of interacting states
\cite{Crisanti,Leuzzi}.
The previous discussion 
suggests 
that 
the 
randomness 
is 
responsible  for producing naturally inverse freezing 
in the GS model
as compared
with 
the 
inverse melting case in the 
BC
model.
In fact, 
to be more precise, it seems that the 
presence 
of  
non-trivial  form of 
randomness,
which means frustration,
\cite{Fradkin} 
would be the  
essential
ingredient
which allows
inverse freezing 
in the GS model
with no need of entropic advantage.
That sets some questions: 
What is the 
actual 
role of 
non-trivial randomness to produce naturally inverse freezing? 
Is 
this natural inverse freezing 
robust
when the level
of frustration is diminished?  
Can
models 
with 
trivial 
randomness
present naturally inverse transitions?

The purpose of the present work is to investigate 
what is the actual role 
of frustration
as  
a basic condition  
to produce naturally 
inverse freezing.  
 That could be achieved by comparing 
two distinct situations, the trivial randomness and the highly frustrated regime
and also
studying
how  
natural 
inverse freezing
would be affected when 
the level of 
frustration is varied.
Therefore, 
the problem 
is not only to find a 
random 
magnetic 
model which presents naturally inverse freezing, but 
also  to get 
one which allows the level of frustration  to vary.
Quite recently, the fermionic Ising spin-glass (FISG) model \cite{Magal1} has 
been proposed as a model able to  
display naturally inverse freezing in a phase diagram temperature {\it versus} the chemical potential $\mu$.
This model has the spin operators $\hat{S}_{i}^{z}$
given as bilinear combinations of creation and destruction 
fermionic
operators which have four eigenstates, two of them non-magnetic \cite{Oppermannnuclear,Alba2002,Wiethege}.  The spin-spin 
coupling $J_{ij}$, likewise the GS model, is a random variable which follows a Gaussian distribution.
Indeed,
the existence of inverse freezing in this model 
could be
expected since there is a  close relationship between
the GS and the FISG models \cite{Castillo}.
For instance, the partition function of these two models 
can be related by a mapping between the anisotropy constant $D$ of the GS model
and 
$\mu$ \cite{Feldmann}.

The FISG model 
is 
also
useful
to examine  how
quantum effects, 
included by the presence of a transverse field $\Gamma$ \cite{Alba2002,Zimmerprb}
can affect  
inverse freezing. 
In Ref. \onlinecite{Magal1}, it has been shown that,  
when $\Gamma$ is increased,
the reentrance in the  $SG/PM$
first order boundary phase 
in 
the phase diagram  
temperature $T$  {\it versus} the chemical potential $\mu$
gradually disappears. 
 This scenario also
suggests 
that $\Gamma$ 
in the FISG model 
plays 
the opposite role of the rather artificial parameter $r$ in the BC model
discussed in Ref. \onlinecite{Schupper1}.
Nevertheless,
the FISG model, in its original form with Gaussian random couplings, is 
a strongly 
frustrated model. 
Therefore, it is not possible to investigate in this model 
how 
the  change of the level of frustration  could affect 
the 
inverse freezing.

An alternative route to accomplish 
the previously mentioned
investigation
would be to use
the FISG model in a version
in which one could 
adjust the level of frustration. 
In that sense, 
the  classical
Hopfield spin glass \cite{Amit1,Amit2} 
can be
quite useful. 
In this model, the interactions $J_{ij}$ between classical Ising spins are given 
below: 
\begin{equation}
J_{i_{}j_{}}=\frac{J}{2N}\sum_{\varrho=1}^{p}\xi_{i}^{\varrho}\xi_{j}^{\varrho}
\label{hebb}~,
\end{equation}
where $\xi_i^{\varrho}=\pm 1$ ($i$ or $j=1,2,...,N$, $N$ is the number of sites) are independent random distributed variables.
This 
type
of 
interaction has been intensively used to study complex systems \cite{Amit1}. 
There are clearly two extreme cases for the classical 
Hopfield Ising spin glass
model.
When $p=1$, 
it becomes
the 
Mattis model \cite{Mattis}, which is a well known 
example 
of trivial 
randomness
\cite{Fradkin}. The second one is when $p\rightarrow \infty$.
In that limit, the thermodynamics
corresponds
to the 
strongly 
frustrated 
regime 
given by a random Gaussian distributed $J_{ij}$ \cite{Provost}.
In particular, 
the replica symmetric mean field solution of the classical 
Hopfield spin glass model \cite{Amit2} can provide a  
useful method 
for the purposes of
the present work. It is described 
in terms of two order parameters, the usual spin 
glass 
$q$
and 
$m_{\varrho}=\frac{1}{N}\sum_{i=1}^{N}\xi_{i}^{\varrho}<S_{i}>$ 
which 
indicates the presence of 
Mattis states that have 
a 
thermodynamics similar to 
the usual ferromagnetism \cite{Amit1,Amit2}. Most important, there is a 
parameter called here degree of frustration, $a=p/N$ which allows  controlling the level of frustration in the theory. It is important to remark that there is a particular 
value $a_{c}$ which for $a>a_{c}$ the effects of frustration are dominant \cite{Amit1,Amit2}. 
As a consequence, 
it is possible 
to interpolate the thermodynamics
from 
the 
trivial 
randomness ($a=0$)
to  
the
strongly 
frustrated 
regime ($a>a_{c}$).

Therefore,
to answer the questions arisen previously, 
we use 
the Hopfield Fermionic 
Ising Spin Glass (HFISG) model.  
In this case, the  spin operators $\hat{S}^{z}_{i}$ are  
defined 
as the 
FISG model, but  
the random spin coupling $J_{ij}$ is given as Eq. (\ref{hebb}). 
The partition 
function is obtained in the functional integral formalism  
using Grassmann fields where
the disorder 
is 
treated with the replica 
method \cite{SK}.
The problem can be reduced to a one-site problem 
by using a similar procedure 
to solve the classical Hopfield spin glass \cite{Amit1,Amit2} 
within
the static approximation (SA) \cite{Bray}  
and 
one-step replica symmetry breaking (1S-RSB) \cite{Parisi}. 
It should be remarked that 
in the HFISG model, as the FISG one \cite{Oppermannnuclear,Wiethege,AlbaMercedes,Zimmerprb},
the replica diagonal component of 
the SG order parameter $q_{\alpha\alpha}$ 
appears as an 
additional order parameter to be solved with the others. 
 Particularly,
this order parameter represents the spin self-interaction that has an imaginary
time dependence for quantum SG models \cite{Bray,Huse}. However, the spin
operators $\hat{S}^{z}_i$
commute with the Hamiltonian operators for the present HFISG model.
Consequently, the $q_{\alpha\alpha}$ has no dynamic, which means that the SA is
exact in the present work\cite{Oppermannnuclear,Alba2002}.

One
important
point  in the present work
is how to locate  first order boundaries phases 
found  
for different degrees of frustration. 
The criterion adopted here follows closely that one suggested in Ref. \onlinecite{DaCostaSalinas} for the classical 
$GS$
model. 
Hence,
we
selected,  
from the set of spin glass solutions in the transition, that one 
which 
meets continuously with the spin glass solution for small 
value of the chemical potential $\mu$. 
In fact, that corresponds to the largest spin glass order parameter which gives the lowest Grand Canonical Potential (see also the discussion in  Ref. \onlinecite{Mottishaw}).
Then, by equating 
the Grand Canonical Potential of the spin glass and paramagnetic solutions, the first order boundary phase is located.

The use of  the 1S-RSB scheme 
also deserves 
some remarks
since our
main interest 
is to obtain boundaries of phase transitions. 
 It has been found in previous works
\cite{Crisanti,Leuzzi,Magal1}
that the use of the replica breaking symmetry (RSB) schemes in GS or FISG models essentially preserves the reentrance in the  $SG/PM$  first order boundary phase associated to the inverse freezing. 
There are only small differences which appear mainly at very low temperatures. 
Even so,  for the HFISG model, we decided to use 1S-RSB scheme 
to check for intermedianted values of
$a$ whether or not  
these differences will remain unimportant.
Moreover, we also analyzed the stability of the RS solutions 
for  a sake of completeness of the work.

This paper is structured as follow: in Section 2, we derived the thermodynamics and  the set of coupled equations for the saddle point order parameters. 
In Section 3, phase diagrams temperature {\it versus} the chemical potential are presented for several values of $a$.
The entropy behavior as function of temperature and the grand canonical potential as function of chemical potential are also discussed. Finally,  Section 4 is reserved 
to 
conclusions.

\section{Model}
The  Hamiltonian considered here is a Hopfield FISG (HFISG)  model
\begin{equation}
{\hat{ H}}= -\sum_{i_{}j_{}} J_{i_{}j_{}}\hat{S}_{i_{}}^{z} \hat{S}_{j_{}}^{z},
\label{ham}
\end{equation}
where 
 $J_{i_{}j_{}}$ is given in Eq. (\ref{hebb}) and the random  $\xi_i^{\varrho}$ follows the distribution
\begin{equation}
P(\xi_{i})=\frac{1}{2}~\delta_{\xi_{i}^{\varrho},+1}+\frac{1}{2}~\delta_{\xi_{i}^{\varrho},-1}.
\label{gaussian}~
\end{equation}
In this model, $\hat S_{i}^{z}=\frac{1}{2}[\hat{n}_{i\uparrow}-\hat{n}_{i\downarrow}]
$ is the spin operator, with $\hat{n}_{i\sigma}=c_{i\sigma}^{\dagger}c_{i\sigma}$ as the 
number operator, $c_{i\sigma}^{\dagger}~(c_{i\sigma})$ are fermions 
creation (destruction) operators and $\sigma=\uparrow$ or $\downarrow$ 
indicate the spin projections.

The partition function in the grand canonical ensemble is given in the Lagrangian path 
integral formalism  where the spin operators are represented 
as  bilinear combinations of 
anticommuting Grassmann fields ($\phi,~\phi^*$) 
\cite{Negele}
\begin{equation}
\begin{split}
Z\{\mu\}&=\int D(\phi^{*}\phi)
~\mbox{e}^{A\{\mu\} }
\end{split}
\label{eqz0}
\end{equation} 
where
\begin{equation}
\begin{split}
A\{\mu\}&=\int^{\beta}_{0}d\tau \{ \sum_{j,\sigma} \phi^{*}_{j\sigma}(\tau)
[ -\frac{\partial}{\partial\tau}+\mu] \phi_{j\sigma}(\tau) 
\\ &-  H(\phi^{*}(\tau),\phi(\tau)) \}
\end{split}
\end{equation}
where $\mu$ is the chemical potential and $\beta=1/T$. $Z\{\mu\}$ 
is Fourier transformed in time which results in
\begin{equation}
Z\{\mu\}=
\int D(\phi^{*}\phi)
e^{A_{\mu}+A_{SG}}
\label{eqz}
\end{equation} 
with
\begin{equation}
A_{\mu}=\sum_{j,\omega}\sum_{\sigma}\phi_{j\sigma}^{\dag}(\omega)
\left[ i\omega + \beta\mu
\right]\phi_{j\sigma}(\omega)
\label{eqao}
\end{equation}
\begin{equation}
A_{SG}=\sum_{\Omega}\sum_{ij}\beta J_{ij}S_{i}^{z}(\omega^{'})S_{j}^{z}(-\omega^{'})
\label{asg1},
\end{equation}
and
\begin{equation}
S_{i}^{z}(\omega^{'})=\frac{1}{2}\sum_{\omega}\sum_{\sigma } \sigma_{s} \phi_{i\sigma}^{\dag}(\omega+\omega^{'})
\phi_{i\sigma}(\omega)
\label{eqs},
\end{equation}
where $\mu$ is chemical potential,
$\omega=(2m+1)\pi$, $\omega^{'}=2m\pi~(m=0,\pm1,\cdots)$ are the Matsubara's frequencies {and $\sigma_{s}=+(-)$ if $\sigma=\uparrow(\downarrow)$} . In this work, the problem is analyzed within 
SA
which 
considers only the term  when $\omega^{'}=0$ in Eq. (\ref{eqs}) \cite{Alba2002,Bray,AlbaMercedes,ZimmerAF}.

The Grand Canonical Potential 
is obtained by using the replica method:
\begin{equation}
\beta \Omega =
-\lim_{n\rightarrow 0} 1/(nN)(\langle\langle {{Z}}(n)\rangle\rangle_{\xi}-1)
\label{replica}
\end{equation}
where $Z(n)\equiv
{{Z}}^n$ and $\langle\langle ... \rangle\rangle_{\xi}$ means the configurational averaged over $\xi$.
Thus:
\begin{equation}
Z(n)=
\prod_{\alpha=1}^{n}
\int  D(\phi_\alpha^*,\phi_\alpha)
\exp[A_{\mu}^{\alpha} + A_{SG}^{\alpha_{stat}}]
\label{zn1}
\end{equation}
where after using $J_{ij}$ given in Eq. (\ref{hebb}), 
the action $A^{stat}_{SG}$  
can be written 
as  
\begin{equation}
A_{SG}^{\alpha_{stat}}= \frac{\beta J}{2N}\sum_{\varrho=1}^{p}\sum_{\alpha=1}^{n}(\sum_{i}
\xi_{i}^{\varrho}S_{i}^{\alpha})^{2} - \frac{\beta Jp}{2N}\sum_{i}\sum_{\alpha=1}^{n}(S_{i}^{\alpha})^{2}
\label{asg},
\end{equation}
with $\alpha$ denoting the replica index and $S_{i}^{\alpha}\equiv S_{i}^{\alpha}(0)$.

The average over $Z(n)$ {given} in Eq. (\ref{zn1}) is discussed in detail in the Appendix A. 
In the present work, the one-step replica symmetry breaking (1S-RSB) ansatz is adopted, in which the replica matrix $\{Q\}$ and the matrix $\{r\}$ are parametrized as:
\begin{equation}
 \begin{array}{cc} q_{\alpha \beta}=\left\{
\begin{aligned}
 \bar{q} \ \   &\mbox{if}\ \alpha=\beta
\\ q_1 \ \    &\mbox{if}\ |\alpha-\beta|\leq x,
\\ q_0  \ \   &\mbox{otherwise}
\end{aligned}\right.
& r_{\alpha \beta}=\left\{
\begin{aligned}
 \bar{r} \ \   &\mbox{if}\ \alpha=\beta
\\ r_1 \ \    &\mbox{if}\ |\alpha-\beta|\leq x
\\ r_0  \ \  &\mbox{otherwise}
\end{aligned}\right.
\end{array}
\label{1srsb}
\end{equation}
and 
order parameters
$m^{\alpha}_{1}$ are invariant with respect to permutations of replicas: $m^{\alpha}_{1}= m$, where $\alpha=1,\cdots n$.
Therefore, the parametrization (\ref{1srsb}) is used 
in Eqs. (\ref{med1})-(\ref{e25}).
As a consequence, the 1S-RSB Grand Canonical Potential  is found as 
\begin{equation} 
\begin{split}
 &\beta\Omega=-\beta\mu+\frac{\beta^2 J^2 a}{2}\left(\bar{r}\bar{q}-(1-x)r_1q_1-xr_0q_0\right)
\\& 
+\frac{a}{2x}\ln \frac{1-\beta J[\bar{q}-q_1+ x(q_1-q_0)]}{1-\beta J(\bar{q}-q_1)}
\\&
-\frac{1}{2}
 \frac{\beta J a q_0}{1-\beta J[\bar{q}-q_1+ x(q_1-q_0)]}
+\frac{\beta J m^2}{2} +\frac{a}{2}
\\ &
\times\ln [1-\beta J(\bar{q}-q_1)] -\lim_{n\rightarrow 0}\frac{1}{ n}\ln\langle\langle 
\Theta(\{r\},m,\xi) \rangle\rangle_{\xi},
\end{split}
\end{equation} 
where 
\begin{equation}
\begin{split}
 &\Theta(\{r\},m,\xi)=\int \prod_{\alpha}^{n} D(\phi^{*}_\alpha,\phi_\alpha)\\&\times
\exp\left[  \frac{\beta^2J^2a}{2}[(\bar{r}-r_1-1/\beta J)\sum_{\alpha=1}^n(S^{\alpha})^2\right] 
\\&\times\exp \left[ (r_1-r_0)\sum_{l=1}^{n/x}\left(\sum_{\alpha=(l-1)x+1}^{lx}S^\alpha\right)^2
\right] \\&\times\exp\left[ r_0(\sum_{\alpha=1}^n S^{\alpha})^2 
+\beta J \sum_{\alpha=1}^{n} (\xi m S^{\alpha}) \right] 
\end{split}
\end{equation}
with  $a=p/N$. 
The quadratic forms into the function 
$\Theta(\{r\},m,\xi)$ can be 
linearized by Hubbard-Stratonovich transformations 
where new auxiliary fields are introduced in the problem. Therefore, one has:
\begin{equation} 
\begin{split}
&\Theta(\{r\},m,\xi)=
\int  Dz
\left[
\int  Dv
\left(\int  Dw \right. 
\right. \\&\times \int D(\phi^{\ast}\phi)
\left. \left. \exp \sum_{\omega\sigma}\phi_{\sigma}^{\ast}(\omega)
G^{-1}(\omega)
\phi_{\sigma}(\omega)\right)^x \right]^{n/x} ,
\end{split} 
\label{e28}
\end{equation}
with $Dy= \frac{dy e^{-\frac{y^{2}}{2}}}{\sqrt{2\pi}}$ ($y=z,\ v,\ w$) and
\begin{equation} 
G^{-1}(\omega)=g^{-1}(\omega)
+ \bar{h}(z,v,w)
+\beta J (\xi m).
\label{e29}
\end{equation}
The local spin glass component of the random field $\bar{h}(z,v,w)$ is defined by
\begin{equation} 
\begin{split}
\bar{h}(z,v,w)&=\beta J[ \sqrt{ar_0}z+\sqrt{a(r_1-r_0)}v
\\&+\sqrt{a(\bar{r}-r_1-1/\beta J)}w ].
\end{split}
\label{e30}
\end{equation}

The functional integral over the Grassmann variables, 
as well as the  
sum over the Matsubara`s frequencies in Eq. (\ref{e28}), 
can be performed following closely the procedure given in references 
\cite{Alba2002,Zimmerprb}.
 Finally, the Grand Canonical Potential within 1S-RSB approximation can be written as 
\begin{equation} 
\begin{split}
& \beta\Omega=\frac{\beta^2 J^2 a}{2}\left(\bar{r}\bar{q}-(1-x)r_1q_1-xr_0q_0\right)
 +\frac{\beta J m^2}{2} 
\\&-\frac{1}{2} \frac{\beta J a q_0}{1-\beta J[\bar{q}-q_1+ x(q_1-q_0)]}
+\frac{a}{2}\ln [1-\beta J(\bar{q}
\\&-q_1)]+\frac{a}{2x}\ln \frac{1-\beta J[\bar{q}-q_1+ x(q_1-q_0)]}{1-\beta J(\bar{q}-q_1)}
\\&
-\beta\mu-\frac{1}{x}\int Dz\langle\langle\ln \int Dv\left(2K(z,v|\xi)\right)^x\rangle\rangle_{\xi}
\label{eq1}
\end{split}
\end{equation} 
where
\begin{equation}
K(z,v|\xi)=\cosh\beta\mu+ \mbox{e}^{\frac{\beta J a}{2} [\beta J (\bar{r}-r_1)-1]}\cosh\bar{H}(z,v,\xi)
\label{Kzv}
\end{equation}
and
\begin{equation}
\bar{H}(z,v|\xi)=\beta J[\sqrt{ar_0}z+\sqrt{a(r_1-r_0)}v+\xi m].
\label{eq20}
\end{equation}
The set of equations for the order parameters $m$, $q_{0}$, $q_{1}$, $\bar{q}$ and  the block size parameter $x$ can be found from Eq. (\ref{eq1}) using the saddle point
conditions.

In particular, the elements of matrix r are given by
\begin{equation}
 r_0=\frac{q_0}{\{1-\beta J[\bar{q}-q_1+x(q_1-q_0)]\}^2}
\label{r01}
\end{equation}
\begin{equation}
 r_1-r_0=\frac{q_1-q_0}{[1-\beta J(\bar{q}-q_1)]\{1-\beta J[\bar{q}-q_1+x(q_1-q_0)]\}}
\end{equation}
\begin{equation}
 \bar{r}-r_1=\frac{1}{\beta J[1-\beta J(\bar{q}-q_1)]}.
\label{rbarra}
\end{equation}

The average over $\xi$ in the  Grand Canonical Potential can be done using the parity properties of the functions dependent on $z$ and $v$. The entropy  can also be obtained directly from the Grand Canonical potential.

The  stability of the RS solution $q_{\alpha\beta}=q$, $r_{\alpha\beta}=r$ ($\alpha\neq \beta$) is studied  using the 
Almeida-Thouless  analysis, in which  the condition for the stable RS solution ( $\lambda_{at}>0$) is obtained as\cite{Amit3} 
\begin{equation}
\begin{split}
 &\lambda_{at}=\left[ 1 + \beta J(q - \bar{q})\right] ^2 \\ &-
 a (\beta J)^2\int Dz\left[ \frac{e^{u}\cosh H(r)}{K(r,\bar{r})} - \left( \frac{e^{u}
        \sinh H(r)}{K(r,\bar{r})}\right) ^2\right] ^2 
\end{split}
\label{linhaAT1}
\end{equation}
with $K(r,\bar{r})=\cosh(\beta\mu) +  e^{u}\cosh H(r)$ and
\begin{equation}
 u=\frac{1}{2}\beta J a [\beta J(\bar{r} - r) - 1],~~~ H(r)= \beta J\sqrt{a r}z +\beta J m.
\label{linhaAT2}
\end{equation}

\section{Results}
\begin{figure*}[ht]
\begin{center}
 \includegraphics[angle=270,width=14cm]{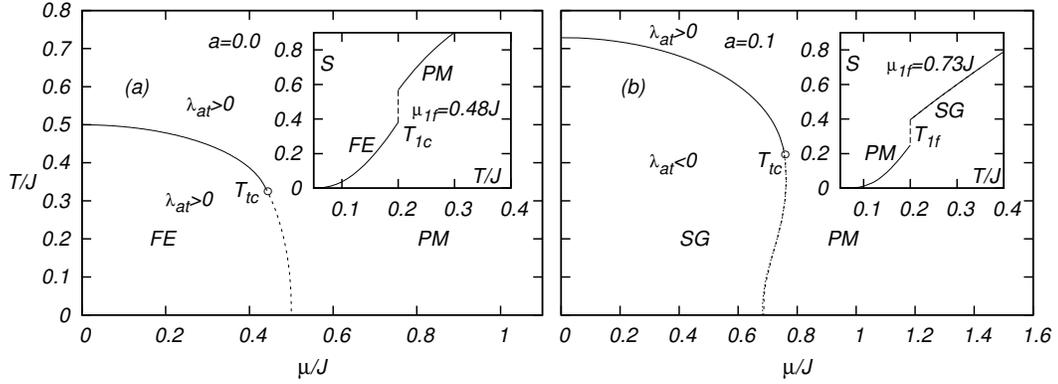}
\end{center}
\caption{Phase Diagrams $T/J$ {\it versus} $\mu/J$ for $a=0.0$ and $a=0.1$. In the phase diagrams, the full lines represent the second order transition and the dashed lines represent the first order transition.  $T_{tc}$ indicates the tricritical point. The insets show the entropy $S$ {\it versus} $T/J$ for specific values of $\mu/J$ in the first order transition, where $T_{1c}$ and $T_{1f}$ represent the first order transition between $FE/PM$ and $SG/PM$ phases, respectively. 
}
\label{fig1}
\end{figure*}

\begin{figure*}[ht]
\begin{center}
 \includegraphics[angle=270,width=14cm]{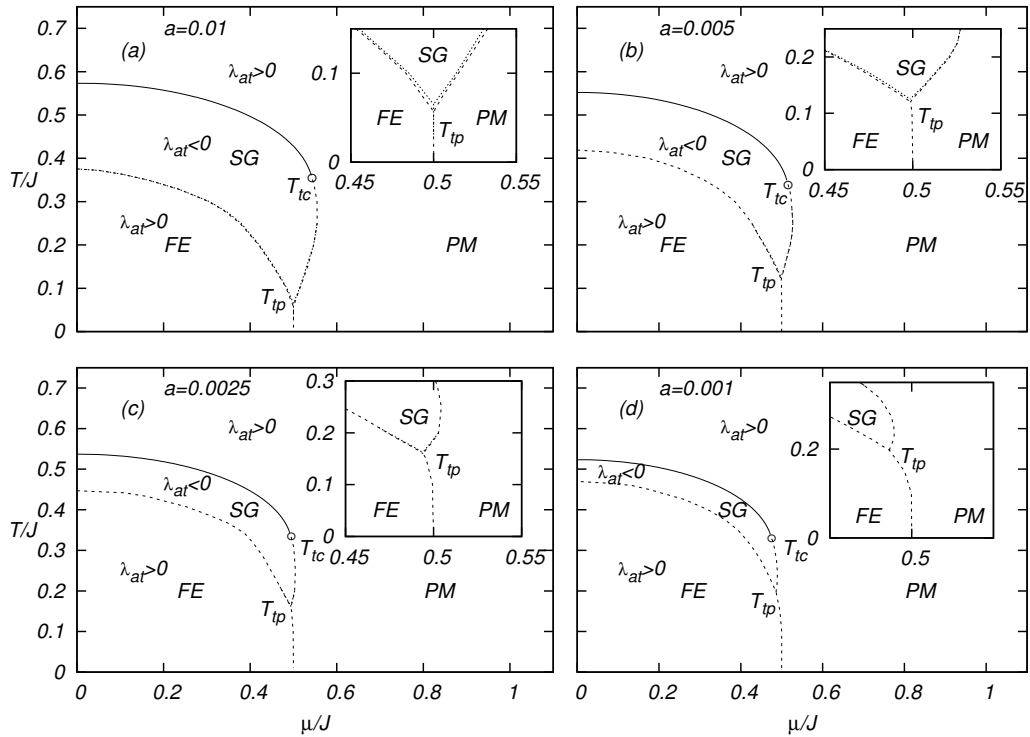}
\end{center}
\caption{Phase diagrams $T/J$ {\it versus} $\mu/J$ for several low values of $a$. $T_{tc}$ is a tricritical point and $T_{tp}$ is a triple point. Dashed lines represent the first order transition and solid lines represent the second order transition. The insets show the location of the triple point in details. Close to the triple point, a small difference between the first order boundary of the 1S-RSB (dotted lines) and RS (dashed lines)  solutions can be seen in the insets.
}
\label{fig2}
\end{figure*}

The numerical solutions for the coupled set for the  saddle point  order parameters $q_{1}$, $q_{0}$, $\bar{q}$ and $m$ 
are displayed in phase diagrams $T/J$ {\it versus} 
$\mu/J$  given below
for several values of the parameter $a$, 
where $T$ is the temperature and  
$J$ is defined in Eq. (\ref{hebb}).  For the numerical  results, $J=1$ is used. 
The 
RS scheme for the order parameters is  also calculated 
when
$q_{0}=q_{1}\equiv q$. 
The stability of such solution is 
investigated by calculating the Almeida-Thouless eigenvalue $\lambda_{AT}$
given in Eqs. (\ref{linhaAT1})-(\ref{linhaAT2}).
Such analysis shows that the $PM$
solutions
and Mattis states ($FE$) 
present $\lambda_{AT}>0$ while 
the $SG$ ones have $\lambda_{AT}<0$. This result could indicate that it is necessary to use  
RSB
schemes 
to locate more adequately the $SG/PM$ first order boundary phase. 
However,  
the results shown in Figs. (\ref{fig1})-(\ref{fig2}) 
have found 
that, in such first order boundary phase, the reentrance
and, therefore, the inverse freezing are not essentially affected by the use of RS or 1S-RSB schemes similar  to the FISG model \cite{Magal1}.

In Fig. (\ref{fig1}), it is presented 
phase diagrams 
which illustrate
two quite distinct situations concerning the degree of frustration.  
The first one ($a=0$), which is shown in Fig. (\ref{fig1}-a), corresponds to 
trivial randomness 
\cite{Fradkin}, 
which means that 
there are no effects of frustration. In this phase diagram, 
for low $T/J$ and small $\mu/J$,  
one gets $FE$ as solution.
For higher $T/J$ 
and/or
larger $\mu/J$, it is found the 
$PM$
solution. 
For small $\mu/J$,
the  
boundary 
phase transition 
between 
$FE$ and $PM$ phase (called here $T_{2c}(\mu)$)
is second order. 
However, when $\mu/J$ increases, it appears a tricritical point located at 
$T_{tc}=J/3$ and $\mu_{tc}=0.438J$ (see Appendix B).
Most important, there is no reentrance in the subsequent first order part of the boundary phase transition
$T_{1c}(\mu)$.
In Fig. (\ref{fig1}-b), the degree of frustration is 
increased 
($a=0.1$) and, for that case,  
effects of frustration become dominant. 
Thus, the 
$FE$ solution is replaced by a $SG$ one
in which, $m=0$, $\bar{q}\neq 0$, $q_{0}\neq 0$ and $q_{1}\neq 0$,  while the $PM$ phase has $m=0$, $\bar{q}\neq 0$, $q_{0}=0$ and $q_{1}=0$.
The freezing temperature 
has also a second order part $T_{2f}(\mu)$.
As for the trivial randomness case,  
a tricritical point  appears, with value $T_{tc}\approx 0.438J$ and $\mu_{tc}\approx 0.752 J$ (see Appendix B).
Nevertheless, 
the 
first order part of the freezing temperature 
$T_{1f}(\mu)$ now
shows a reentrance 
which indicates
the existence of inverse freezing, as it can be seen in the insert in Fig. (\ref{fig1}-b), 
which displays
the entropy as function of $T/J$. There, 
it is 
shown that entropy of the $SG$ phase is larger than the $PM$ one 
for that particular degree of frustration.  It should be remarked that this 
phase diagram
is 
quite similar  to that one found for 
the FISG model \cite{Magal1}. The location of the boundaries phases in Fig. (\ref{fig1}) has  been checked in the  limits $\mu=0$ and $T=0$ as it can be seen in Appendices B and C.

There is a more complex scenario 
as compared with that one
described in Fig. (\ref{fig1}-b) when $a\rightarrow 0$.
In Fig. (\ref{fig2}-a)
($a=0.01$), 
the solution for the order  parameters shows a phase diagram  which illustrates this new scenario. For a small $\mu/J$, 
when $T/J$ is decreased, there is a second order phase transition between the $PM$ and $SG$ phases. 
However, for even lower temperatures, there is another phase transition, which is now a first order one, between the $SG$  phase and $FE$ region, which is
given now by $m\neq 0$, $\bar{q}\neq 0$, $q_{0}\neq 0$ and $q_{1}\neq 0$.
For this particular value of $a$, 
$SG$ and $FE$  solutions occupy   
approximately equal sizes in the phase diagram.
The 
freezing
temperature has a similar 
behavior 
to that one found in Fig. (\ref{fig1}-b).  
It has
a second order part $T_{2f}(\mu)$ for small $\mu$, then it appears a tricritical point at $T_{tc}\approx 0.366J$ and $\mu_{tc}\approx 0.539J$  (see Appendix B). Below this point, $T_{1f}(\mu)$ presents a reentrance which allows, for an adequate constant $\mu$, crossing from the $SG$ phase to the $PM$ one when the temperature is decreasing.
Nonetheless,  this first order boundary phase has a complex nature.
It appears 
a triple point at ($T_{tp}$,$\mu_{tp}$)  where
$FE$, $PM$ and $SG$ 
phases coexist.  
Below this point,
the first order boundary phase $T_{1c}(\mu)$ diplays no
reentrance, 
as in the case $a=0$.

For smaller values of
$a$,  
as shown in Figs. (\ref{fig2}-b)-(\ref{fig2}-d), the region 
where $FE$ solutions are found  becomes increasingly larger than the $SG$ one  
which is consistent with earlier results found in the classical Hopfield spin glass model which displays, in a phase diagram $T$ {\it versus} $a$,  a dominance of the Mattis states
when $a \rightarrow$ 0\cite{Amit1}.
Even so, in this new situation,  
the location 
$T_{2f}(\mu)$ is not affected  so much.  However, the tricritical point is displaced for smaller and lower values of $\mu/J$ and $T/J$, respectively. 
In comparison,
the triple point 
is 
displaced for smaller values
of $\mu/J$ 
and higher values of $T/J$. 
As a consequence, $T_{1f}(\mu)$ and $T_{1c}(\mu)$
appear 
in a decreasing and increasing  
range of temperature, respectively. 
Nevertheless, most important,
the reentrance in $T_{1f}((\mu)$ 
is gradually 
suppressed 
when $a\rightarrow 0$. 
However, the insert in Fig {(\ref{fig2}-d)} shows that even when the level of frustration is very weak,  
and $T_{2f}(\mu)$ appears in a very short range of temperature,  a reentrance in such first order boundary phase  is still preserved.

\begin{figure}[htb]
\begin{center}
 \includegraphics[width=8.66cm]{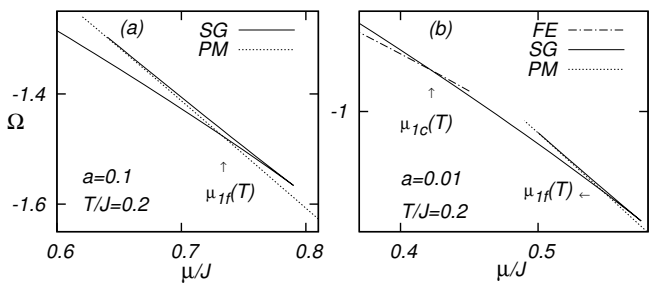}
\end{center}
\caption{Grand canonical potential $\Omega$ {\it versus} $\mu/J$ for $T/J=0.2$ and two values of $a=0.1$. The full, dotted and dashed-dotted lines represent the $\Omega$ of SG, PM and FE solutions, respectively. 
}
\label{fig3}
\end{figure}

\begin{figure}[htb]
\begin{center}
 \includegraphics[width=8.6cm]{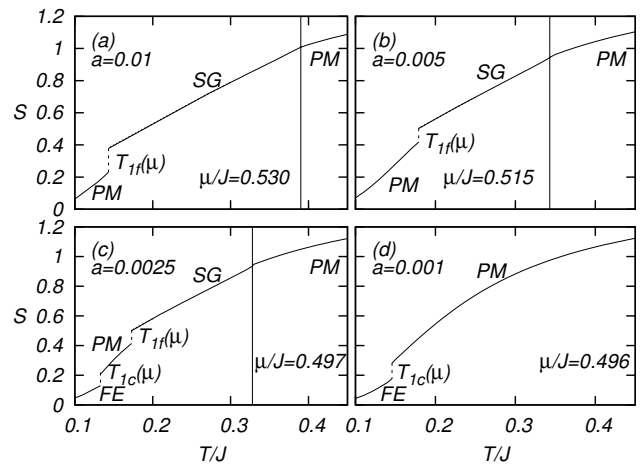}
\end{center}
\caption{Entropy $S$ {\it versus} $T/J$ for several values of $a$ and $\mu/J$. The labels $T_{1c}(\mu)$ and $T_{1f}(\mu)$ indicate the first order transition between the  $FE/PM$ and $SG/PM$ phases, respectively. The full vertical line indicates the second order transition between $SG/PM$ phases. 
}
\label{fig4}
\end{figure}

The procedure to locate the first order boundary lines  in the previous phase diagrams  
is illustrated in fig. (3), where the grand canonical potential {\it versus} $\mu/J$  is plotted for $T/J=0.2$. In fig. (3),  the set of multiple $SG$ solutions is represented by branches of full lines,
where the chosen $SG$ solution is that one which meets continuously with the only one $SG$ solution  available for small values of $\mu/J$. From that $SG$ solution, the first order boundary phase is obtained by equating the grand canonical potential of SG, PM and FE solutions.

The  corresponding
behavior of the 
entropy as a function of $T/J$ for the values of $a$ used in Fig. (\ref{fig2}) is shown in 
Fig. (\ref{fig4}). This figure 
illustrates
the gradual suppression of inverse freezing when $a\rightarrow 0$.  
For instance, in Fig. (\ref{fig4}-a), the value of $\mu=0.53J$ is chosen to cross the 
reentrant first order boundary phase
in $T_{1f}(\mu)$ 
in Fig. (\ref{fig2}-a). The figure shows that the entropy of $SG$ phase is 
larger
than the entropy of the $PM$ 
one
in the first order transition which, as in the inset of Fig. (\ref{fig1}-b), indicates the existence of inverse freezing.
The same procedure has been adopted in Figs. (\ref{fig2}-b)-(\ref{fig2}-d), values of $\mu$ are adjusted to be close to $\mu_{tp}$. 
The result found   
in Fig. (\ref{fig2}-b) is similar to that one found in Fig.(\ref{fig2}-a). However,  Fig. (\ref{fig2}-c) displays the entropy behavior when $T_{1f}(\mu)$ and $T_{1c}(\mu)$  are now crossed. 
The first crossing is 
in
the 
reentrant line transition $T_{1f}(\mu)$
giving an inverse freezing as in Figs. (\ref{fig2}-a)-(\ref{fig2}-b). 
The second one in $T_{1c}(\mu)$ gives a usual phase transition in which 
the $PM$ phase is more entropic than  
the $FE$ one, as already found in Fig. (\ref{fig1}-a). In Fig. (\ref{fig2}-d) the reentrance in $T_{1f}(\mu)$ is 
almost suppressed. Thus,  for the chosen value of $\mu$, there is only one crossing in $T_{1c}(\mu)$  which gives the entropy behavior of a usual phase transition as that one found in Fig. 
(\ref{fig1}-a).

\section{Conclusion}

In the present work, it has been studied 
the HFISG
model 
in which the random spin-spin coupling $J_{ij}$, instead of the usual Gaussian distribution,
is given  
by Eq. (\ref{hebb}). 
For this particular choice of $J_{ij}$, the problem can be treated in a mean field framework which allows  adjusting the level of frustration. Therefore, it is possible 
to study not only the existence of a natural inverse freezing in the limits of trivial randomness and strong frustration but also
how changing the level of frustration could affect such transition.
In this approach, 
 within  1S-RSB scheme, 
besides the replica non-diagonal $SG$ orders parameters $q_{0}$, $q_{1}$ and the block size parameter $x$,  there is also 
the replica diagonal order parameter $\bar{q}$. The set of order parameters is completed with
$m$ which corresponds to the presence of Mattis states ($FE$) \cite{Amit1,Amit2}.   The 
coupled equations for $q_{0}$, $q_{1}$, $x$, $\bar{q}$ and $m$ are solved for several values of  degree of frustration $a$ in a phase diagram $T$ {\it versus} chemical potential $\mu$. In particular, the RS solution is also obtained when 
$q_{0}=q_{1}\equiv q$.

The comparison among the several scenarios displayed in Figs. (\ref{fig1})-(\ref{fig2}) 
elucidates
the role of frustration as the 
essential
ingredient responsible  for producing naturally
inverse freezing in the HFISG model. In other words, an inverse freezing without  any need of an 
artificial
entropic advantage.
For instance, the comparison between results given in 
Figs. (\ref{fig1}-a) and  (\ref{fig1}-b) shows that only in the case $a=0.1$ 
(the strong frustrated regime) a 
reentrant first order boundary phase appears. 
The trivial randomness is not able 
to generate naturally a reentrance and, therefore, an inverse transition.
Furthermore, in Fig.  (\ref{fig2}), when the level of frustration is diminished,  
the range in temperature of the $SG/PM$ first order boundary phase is decreased but, most importantly, 
the reentrance in this
first order 
boundary phase 
is gradually suppressed,  what implies 
in the gradual suppression
of inverse freezing 
until the complete disappearing in the trivial randomness case ($a=0$).
Nevertheless, 
whatever the level of frustration, 
there is always a 
reentrance in $SG/PM$ first order boundary phase which gives an inverse freezing. 
In contrast, 
the range of the $FE/PM$ first order boundary phase 
is simply
increased, nothing else happens.  
These previous features lead to the conclusion 
that frustration in any level is 
the 
necessary 
condition 
to create naturally the entropic advantage of the $SG$ phase as compared with the $PM$ one which generates an inverse freezing.  Albeit, these results are restrict to a particular model, we suggest that the role of frustration as an essential ingredient to rise naturally inverse freezing could be more general.

One last remark must be done. 
The present approach could be used directly in GS model with the coupling $J_{ij}$ as given in Eq. (\ref{hebb}) replacing the original Gaussian distributed one \cite{GS}.
However, the great advantage of the HFISG model is that it would also allow 
 studying the present problem 
in the presence of
a
transverse magnetic field $\Gamma$ \cite{Alba2002,Zimmerprb,ZimmerAF}. In Ref. \onlinecite{Magal1}, 
it has been shown that $\Gamma$ tends to destroy the inverse freezing. 
Therefore, the presence of an additional 
$\Gamma$
would lead
to another question: 
what would  happen with the
inverse freezing in the HFISG model
when the level of frustration and the strength of quantum effects are simultaneously 
changed?
In that case, 
it would possible to 
investigate the robustness of the inverse freezing 
by 
adjusting simultaneously the 
degree
of frustration and $\Gamma$.
 This question is currently under investigation.

\section*{Acknowledgments}

SGM  acknowledges the hospitality of the Departamento de Ciencias de la Tierra y la Materia Condensada-Universidad de Cantabria where this work was concluded and the support of Fundacion Car\'olina/Spain. 
This work was partially supported by the Brazilian agencies 
CAPES (Coor\-de\-na\c{c}\~ao de A\-per\-fei\c{c}\-oamento de Pe\-sso\-al de
Ni\-vel Su\-pe\-rior) and CNPq (Conselho Nacional de Pesquisas Cient\'ificas).

\appendix 
\section{The average over $Z(n)$ }\label{apendicea}

In this appendix,  the averaging procedure of the partition function of the HFISG model 
 is  introduced following closely Ref. \onlinecite{Amit1}. 
 The first term in the action $A^{stat}_{SG}$ (see Eq. (\ref{asg})) can be linearized by a Hubbard-Stratonovich transformation by introducing $n \times p$ 
auxiliary fields 
$m_{\varrho}^{\alpha}$ which are splitted in two subsets with $n \times (p-l)$ and $n \times l$ terms. 
Therefore,
 \begin{equation}
\begin{split}
&\exp(A_{SG}^{stat})=\exp\left[ -\frac{\beta Jp}{2N}\sum_{\alpha}\sum_{i}(S_{i}^{\alpha})^{2}\right]
\\ &\times 
\int_{-\infty}^{\infty} D m_{\nu}^{\alpha}  
\exp\left\lbrace \tau \sum_{\nu=1}^{l}\sum_{\alpha}\left[ -\frac{1}{2}(m_{\nu}^{\alpha})^2 + \eta_{i,\alpha}^{\nu}
\right] \right\rbrace \\ &\times
\int_{-\infty}^{\infty} D m_{\varrho}^{\alpha}  \exp\left\lbrace \tau \sum_{\varrho=l+1}^{p}\sum_{\alpha}\left[ -\frac{1}{2}(m_{\varrho}^{\alpha})^2  + \eta_{i,\alpha}^{\varrho}
\right] \right\rbrace,
\label{mu}
\end{split}
\end{equation}
where $\tau=\beta J N$,  $\eta_{i,\alpha}^{\nu(\varrho)}=\frac{1}{N}\sum_{i}\xi_{i}^{\nu(\varrho)}S_{i}^{\alpha}m_{\nu(\varrho)}^{\alpha}$  and $D m_{\nu(\varrho)}^{\alpha}=\prod_{\nu(\varrho)}\prod_{\alpha}\frac{dm_{\nu(\varrho)}^{\alpha}}{\sqrt{2\pi}}$.

It is assumed that the relevant contributions come from $m_{\nu}^{\alpha}$ which are order unity, while $m_{\varrho}^{\alpha}$ is of order $1/\sqrt{N}$. Therefore, the average over the $p - l$ independent random variables $\xi_{i}^{\varrho}$ can be done using $P(\xi_{i}^{\varrho})$ given in Eq. (\ref{gaussian}) which results in:

\begin{eqnarray}
\langle\langle \exp\left[ \beta J\sum_{\varrho=l+1}^{p}\sum_{\alpha}(\sum_{i}\xi_{i}^{\varrho}S_{i}^{\alpha})m_{\varrho}^{\alpha}\right] \rangle\rangle_{\xi}= \nonumber \\ \exp\left[ \sum_{i}\sum_{\varrho=s}^{p} \ln\left( \cosh (\beta J \sum_{\alpha}S_{i}^{\alpha}m_{\varrho}^{\alpha})\right) \right]. 
\label{mu1}
\end{eqnarray}

The argument of the exponential in the right hand side of Eq. (\ref{mu1}) can be expanded up to second order in $m_{\varrho}^{\alpha}$. The result is a quadratic term of the spins variables $S_{i}^{\alpha}$ in the last exponential of Eq. (\ref{mu}). This term can be linearized by introducing the spin glass order parameter $q_{\alpha\beta}$ using the integral representation of the delta function as
\begin{equation}
\begin{split}
&\int_{-\infty}^{\infty}\frac{dr_{\alpha\beta}^{'}}{2\pi} \exp\left[ i r_{\alpha\beta}^{'}(q_{\alpha\beta}-\frac{1}{N}\sum_{i}S_{i}^{\alpha}S_{i}^{\beta})\right]\\ &= \delta(q_{\alpha\beta}-\frac{1}{N}\sum_{i}S_{i}^{\alpha}S_{i}^{\beta}). 
\label{mu2}
\end{split}
\end{equation}

Therefore, the exponential involving $m_{\varrho}^{\alpha}$ in Eq. (\ref{mu}) can be written as: 
\begin{equation}
\begin{split}
&\exp\left\{\beta N
\sum_{\varrho=l+1}^{p}\sum_{\alpha}\left[-\frac{1}{2}(m_{\varrho}^{\alpha})^2
+\frac{1}{N}(\sum_{i}\xi_{i}^{\varrho}S_{i}^{\alpha})m_{\varrho}^{\alpha}\right]\right\}\\&=
\int_{-\infty}^{\infty} 
\prod_{\alpha\beta} \frac{dq_{\alpha\beta}
d\tilde{r}_{\alpha\beta}}{2\pi} 
\exp \left[  \frac{\beta}{2} \sum_{\varrho=l+1}^{p} m_{\varrho}^{\alpha}\Lambda_{\alpha\beta} 
m_{\varrho}^{\beta} D_{\alpha\beta} \right] ,
\label{e20}
\end{split}
\end{equation}
with
\begin{equation}
D_{\alpha\beta}=i\sum_{\alpha\beta}{\tilde{r}}_{\alpha\beta}(q_{\alpha\beta}-\frac{1}{N}\sum_{i}S_{i}^{\alpha}
S_{i}^{\beta})
\end{equation}
where the matrix element
\begin{equation}
\Lambda_{\alpha\beta}=(1- \beta q_{\alpha\alpha})\delta_{\alpha\beta} + 
\beta q_{\alpha\beta} (1- \delta_{\alpha\beta})
\label{e21}~.
\end{equation}
Introducing Eqs. (\ref{e20})-(\ref{e21}) into Eq. (\ref{mu}), the
$m_{\varrho}^{\alpha}$ fields can be integrated to give:
%
\begin{equation}
\begin{split}
&\langle\langle \exp (A_{SG}^{stat}) \rangle\rangle_{\xi} =
\exp\left[
-\frac{\beta\ J\ p}{2N}\sum_{i}(S_{i}^{\alpha})^{2} \right]
\\&\times
\langle\langle \int^{+\infty}_{-\infty} Dm_{\nu}^{\alpha} 
\exp\left\{\tau \sum_{\nu=1}^{l}\sum_{\alpha} 
\left[ -\frac{1}{2}(m_{\nu}^{\alpha})^{2}+
\eta_{i,\alpha}^{\nu}
\right]\right\}
\rangle\rangle_{\xi} 
 \\ &\times
\int^{\infty}_{-\infty}\prod_{\alpha\beta}
\frac{dq_{\alpha\beta}{\tilde{r}}_{\alpha\beta}}{2\pi}
\exp\{ D_{\alpha\beta}-
\frac{1}{2}(p-l)Tr\ln\underline{\underline{\Lambda}}) \}.
\label{e22}
\end{split}
\end{equation}

Assuming $l=1$ in Eq. (\ref{e22}), the averaged partition function is given as

\begin{equation}
\begin{split}
&\langle\langle Z(n) \rangle\rangle_{\xi}=
\int^{\infty}_{-\infty}Dm^{\alpha}_{1}\int^{\infty}_{-\infty}\prod_{\alpha\neq\beta}
\frac{dq_{\alpha\beta}d{\tilde{r}}_{\alpha\beta}}{2\pi}
\prod_{\alpha}\frac{dq_{\alpha\alpha} d\tilde{r}_{\alpha\alpha}}{2\pi} 
\\ &\times \exp\left[ i\sum_{\alpha}{\tilde{r}}_{\alpha\alpha}q_{\alpha\alpha}+i\sum_{\alpha\neq\beta}
\tilde{r}_{\alpha\beta}q_{\alpha\beta} -\frac{\beta JN}{2}\sum_{\alpha}(m^{\alpha}_{1})^{2} \right] 
\\ &\times \exp \left[ -\frac{p-1}{2} 
Tr\ln\underline{\underline{\Lambda}} \right] \langle\langle
\Theta(\tilde{r}_{\alpha\beta},\tilde{r}_{\alpha\alpha},m^{\alpha}_{1})\rangle\rangle_{\xi}
\end{split}
\label{med1}
\end{equation}
where the matrix $\underline{\underline{\Lambda}}$ is defined in Eq. (\ref{e21}) and
\begin{equation}
\begin{split} 
&\Theta(\tilde{r}_{\alpha\beta},\tilde{r}_{\alpha\alpha},m^{\alpha}_{1})
=\prod_{\alpha}^{n}\int D(\phi^{\ast}_{\alpha}\phi_{\alpha}) 
\exp\left[  A^{\alpha_{0}}_{i} \right.
\\& 
\left. +\sum_{j}\sum_{\sigma\alpha}\sum_{\omega}
\phi_{j\sigma\alpha}^{\ast}(\omega)
g_{j}^{-1}(\omega) \phi_{j\sigma\alpha}(\omega)\right] 
\label{e25}
\end{split}
\end{equation}
with 
\begin{equation}
\begin{split}
 A^{\alpha_{0}}_{i}&= 
-i\sum_{\alpha\neq\beta}{\tilde{r}}_{\alpha\beta}(\frac{1}{N}
\sum_{i}S_{i}^{\alpha} S_{i}^{\beta}) 
-\sum_{\alpha}(\frac{\beta J p}{2}
\\ & +i{\tilde{r}}_{\alpha\alpha})
 \frac{1}{N}\sum_{i}
(S_{i}^{\alpha})^{2}
+\beta J
\sum_{\alpha} (\sum_{i}\xi_{i}^{1}S_{i}^{\alpha})
m_{1}^{\alpha}
\end{split}
\end{equation}
and $g_{j}^{-1}(\omega)= i\omega + \beta\mu
$. The trace of the matrix $\underline{\underline{\Lambda}}$ given in  Eq. ({\ref{e21}})
is
obtained 
in terms of its eigenvalues.

The Grand Canonical Potential is found introducing Eq. (\ref{med1}) in Eq. (\ref{replica}) which
is evaluated at the saddle point.
Thus,
\begin{equation} 
-i{\tilde{r}}_{\alpha\alpha}=\frac{\beta^{2}J^{2}}{2}\langle(m^{\alpha}_{1})^{2}\rangle=\frac{\beta^{2}J^{2}}{2}p\ r_{\alpha\alpha}  
\label{e27}
\end{equation}
and
\begin{equation} 
-i{\tilde{r}}_{\alpha\beta}=\frac{\beta^{2}J^{2}}{2}\langle (m^{\alpha}_{1}m^{\beta}_{1})\rangle=
\frac{\beta^{2}J^{2}}{2}p\ r_{\alpha\beta};~~~\alpha \neq \beta. 
\label{e27d}
\end{equation}

\section{Tricritical Points}\label{apendiceb}

This appendix presents a procedure to obtain the tricritical point $(\mu_{tc}, T_{tc})$ for both cases: 
$a=0$ and $a> 0$.
For $a=0$, the Landau expansion of the Grand Canonical Potential, Eq. (\ref{eq1}),
in powers of $m$ gives

\begin{equation}
\beta\Omega(m)=A_{0} + A_{2} m^{2} + A_{4} m^{4} + A_{6} m^{6}+ ...
\end{equation}
with
\begin{equation}
A_{2} = \beta J [ 1 - (\beta J)/(1 + \cosh(\beta \mu))]/2, 
\end{equation}
\begin{equation}
A_{4} = \beta^{4} J^4 [2 - \cosh (\beta \mu)] sech [(\beta \mu)/2]^4/96.
\end{equation}
The tricritical point is obtained when $A_{2}=A_{4}=0$.  The position of $T_{2c}$ in the Fig. (\ref{fig1}-a) for $\mu=0$ can be also checked from 
the equation $A_{2}=0$.

For $a >0$, the $PM/SG$ phase transition is investigated. In this case, it is assumed that there is no essential difference between RS and 1S-RSB schemes concerning the location of the tricritical point.\cite{Crisanti}
Therefore, we start with the thermodynamic potential within the RS solution ($q\equiv q_1=q_0$, $r\equiv r_1=r_0$ and $x=0$) written explicitly as a function of the order parameters $r$ and $\bar{r}$, and with $m=0$:
\begin{equation}
\begin{split} 
 \beta \Omega= -\beta \mu + \frac{a}{2}[\beta J \bar{r}-\frac{r}{r-\bar{r}}-\ln{(\bar{r}-r)}]
- \int Dz
\\
\times \ln [\cosh \beta\mu +\mbox{e}^{[\frac{\beta^2 J^2 a}{2}(\bar{r}-r)-\frac{\beta J a}{2}]} \cosh\beta J \sqrt{ar}z]
\label{aa1}
\end{split}
\end{equation}
 where $r$ and $\bar{r}$ are given by Eqs. (\ref{r01}) and (\ref{rbarra}) within the RS solution, respectively. 

Equation (\ref{aa1}) is expressed as  an expansion in powers of $r$, which is related to the $SG$ order parameter.
Therefore,
\begin{equation}
 \beta \Omega= \sum_{i=1}^{4}f_i(\bar{r},\mu,T,a)r^i
\label{omegae}
\end{equation}
where $\bar{r}(r,\mu,T,a)$ is obtained by a saddle point solution of $\Omega$. In this case, $\bar{r}(r,\mu,T,a)$ can also be written in the form of a series
\begin{equation}
 \bar{r}=\bar{r}_0+\bar{r}_1 r +\bar{r}_2 r^2
\label{rbr}
\end{equation}
with $r_1=0$  as a result,
\begin{align}
 \bar{r}_0=[\beta J(1-\beta J X_0)]^{-1},
\\
 \bar{r}_2=-\frac{2+\beta^6 J^6 a^2 \bar{r}_0^3 X_0^2(1-X_0)}{\bar{r}_0[2-\beta^4 J^4 a\bar{r}_0^2 X_0(1-X_0)]},
\end{align}
and 
\begin{equation}
X_0=\frac{\exp{(\frac{\beta^2 J^2 a \bar{r}_0}{2})}}{\mbox{e}^{\beta J a/2} \cosh\beta\mu +\exp{(\frac{\beta^2 J^2 a \bar{r}_0}{2})}}.
\end{equation}

Now, Eq. (\ref{rbr}) is introduced into the coefficients of expression (\ref{omegae}),  which are expanded in powers of $r$  again. The resulting expression is then expressed in powers of the $SG$ order parameter $q$ by expanding $r$:
\begin{equation}
\beta \Omega= F_0 +\frac{F_2 \ q^2}{(1-\beta J \bar{q}_0)^4} +\frac{F_3 \ q^3}{(1-\beta J \bar{q}_0)^6}
+\frac{F_4\ q^4}{(1-\beta J \bar{q}_0)^8} 
\label{landau}
\end{equation} 
with
\begin{align}
 F_2=\frac{a}{4}(\beta^4 J^4 a X_0^2-\frac{1}{\bar{r}_0^2})
\\
 F_3=\frac{a}{3}(\beta^6 J^6 a^ 2 X_0^3+\frac{1}{\bar{r}_0^3})
\end{align}
\begin{equation}
\begin{split}
F_4=\frac{a}{4}\{\frac{\beta^4 J^4 a^ 2 \bar{r}_2}{2} (X_0-X_0^2)(\bar{r}_2+2\beta^2 J^2 a X_0)
\\
+\frac{\beta^8 J^8 a^3 X_0^2}{12}(45 X_0^2-12 X_0+1)\}
\\
-\frac{1}{\bar{r}_0^4}[(\bar{r}_0\bar{r}_2-1)^2+\frac{1}{2}]
-6\beta J (1-\beta J \bar{q}_0)F_3
\end{split}
\end{equation}
where $\bar{q}_0=1/[\beta J(1+\sqrt{a})]$.
The $PM/SG$ second order phase transition occurs when $F_2=0$ and $F_4>0$ with the tricritical point located when $F_2=0$ and $F_4=0$. 
{ In particular, one can use the condition $F_2=0$ with Eqs. (B7) and (B9) to obtain the critical temperature $T_{2f}$ by solving}
\begin{equation}
 \cosh{\beta_{2f} \mu}=\exp{(\frac{J\beta_{2f}\sqrt{a}}{2})[J\beta_{2f}(1+\sqrt{a})-1]}
\end{equation}
{ where $\beta_{2f}=1/T_{2f}$. For $\mu=0$, it is recovered the result $T_{2f}=0.729$ for $a=0.1$} 

\section{First Order Transition at Zero Temperature}\label{apendicec}

Here a procedure is presented to obtain the first order boundary phases of $FE/PM$ phases $(a=0)$ and $SG/PM$ phases $(a>a_{c})$ 
at $T=0$ within the RS and 1S-RSB solutions.
 
For $a=0$, the grand canonical potential of FM solution at $T=0$ is
$\Omega_{fm}= - \mu -\frac{J}{2}$.
By comparing $\Omega_{fm}$ with the grand potential potential of PM phase, $\Omega_{pm}=-2\mu$, the first order boundary  is located at $\mu=0.5J$.  For the cases shown in Fig. (\ref{fig2}), it is recovered the grand canonical potential for $a=0$.

For $a>a_{c}$, the effects of frustration are dominants. In this case for RS solution, $\bar{q}-q\varpropto T$ for $T\rightarrow 0$. Therefore, close to the transition, which means $\mu/J>\frac{a\beta J(\bar{q}-q)}{2[1-\beta J(\bar{q}-q)]}$, the  $\Omega_{sg}$ at $T= 0$ is
\begin{equation}
\Omega_{sg}=-2\mu + \sqrt{2ar}\left[y~ erfc\left(\frac{y}{\sqrt{2}} \right) - \frac{\mbox{e}^{-y^{2}/2}}{\sqrt{2\pi }}\right]
\label{omegax}
\end{equation}
where $erfc(z)=1-erf(z)$ ($erf(z)$ is the 
error function), $y\equiv \frac{1}{\sqrt{ a r }}\left(\mu/J - \frac{a}{\sqrt{ 2\pi a r  }  e^{y^{2}/2} - 2}\right)$
and
$r=erfc\left( \frac{y}{\sqrt{2}}\right)$. From the condition $\Omega_{sg}=\Omega_{pm}$, $y$ and $r$ can be solved which allows finding $\mu_{1f}(T=0)$.  Particularly,  $\mu_{1f}(T= 0)$=0.689 for $a=0.1$.

In 1S-RSB scheme, for $T\rightarrow 0$,
$\chi\equiv\beta J(\bar{q}-q_1)$ is independent of $T$ and $x=\gamma T/J$. Therefore,  close to the transition, where $\mu/J>\frac{a\beta J(\bar{q}-q_{1})}{2[1-\beta J(\bar{q}-q_{1})]}$,  the $\Omega_{sg}$ at $T= 0$ is

\begin{equation}
\begin{split}
& \Omega_{sg}= \frac{J a}{2} \left[ \bar{r}\chi  + \frac{q_{1}}{1-\chi} + \gamma(r_{1}q_{1}-r_{0}q_{0})\right.  \\ &- \left.  \frac{q_{0}}{1 - \chi - \gamma(q_{1}-q_{0})} +\frac{1}{\gamma} \mbox{ln}\left(\frac{1-\chi - \gamma(q_1-q_0)}{(1-\chi)}\right) \right] \\&- 2\mu - \frac{J}{\gamma} \int Dz \mbox{ln}\frac{I(z)}{2} 
\end{split}
\label{1srsbsch}
\end{equation}
where
\begin{equation}
\begin{split}
&I(z)= e^{\gamma \vartheta_{+}}\left[ 1 + \mbox{erf}\left(\frac{\eta_{+}}{\sqrt{2 a (r_{1}-r_{0})}}\right) \right] \\ &+ e^{\gamma \vartheta_{-}}\left[ 1 + \mbox{erf}
\left(
\frac{
\eta_{-}
}{
\sqrt{2 a (r_{1}-r_{0})}
}
\right)
 \right] 
\\ &+ \mbox{erf}\left( 
\frac{
\bar{u}+ \sqrt{a r_{0}}z
}{
\sqrt{2 a (r_{1}-r_{0})}
}
\right) + \mbox{erf}\left( 
\frac{
\bar{u} - \sqrt{a r_{0}}z
}{
\sqrt{2 a (r_{1}-r_{0})}
}
\right)  
\end{split}
\label{1rscompl}
\end{equation}
with $\eta_{\pm}=\pm \sqrt{ar_{0}}z + a\gamma(r_1-r_0) -\bar{\mu}$, $\bar{\mu}=\mu/J - \frac{a}{2}\frac{\chi}{1-\chi}$ and $\vartheta_{\pm}=\pm \sqrt{ar_{0}}z +  \frac{a}{2}\gamma (r_1-r_0) -\bar{\mu}$.
The previous results for RS solution are recovered from eqs. (\ref{1srsbsch})-(\ref{1rscompl}) when $q\equiv q_{1}\equiv q_{0}$. 
Again, equations for $\chi$, $q_0$, $q_1$, $r_{0}$, $r_{1}$ and $\gamma$ can be obtained from $\Omega_{sg}$ and solved. In this case, $\mu_{1f}(T= 0)$=0.682 for $a=0.1$.

\end{document}